\documentclass[aps,pra,reprint,superscriptaddress]{revtex4-1}
\usepackage{graphicx}
\usepackage{times,epsfig,amssymb,amsfonts,amsmath}
\usepackage{bm}
\usepackage{epstopdf}
\usepackage{lipsum}

\begin{document}
\title{Deviations beyond the Kibble-Zurek mechanism in a Spin-Orbit-Coupled Bose-Einstein Condensate with phenomenological damping}

\author{Jun-Hang Ren}
\affiliation{Key Laboratory of Quantum Information, University of Science and Technology of China, Hefei 230026, China}

\author{Sheng Liu}
\email[]{shengliu@ustc.edu.cn}
\affiliation{Key Laboratory of Quantum Information, University of Science and Technology of China, Hefei 230026, China}
\affiliation{Anhui Province key laboratory of Quantum Network, University of Science and Technology of China, Hefei 230026, China}
\affiliation{Hefei National Laboratory, University of Science and Technology of China, Hefei 230088, China}

\author{Yong-Sheng Zhang}
\email[]{yshzhang@ustc.edu.cn}
\affiliation{Key Laboratory of Quantum Information, University of Science and Technology of China, Hefei 230026, China}
\affiliation{Anhui Province key laboratory of Quantum Network, University of Science and Technology of China, Hefei 230026, China}
\affiliation{Hefei National Laboratory, University of Science and Technology of China, Hefei 230088, China}

\date{\today}

\begin{abstract}
We investigate the quench dynamics in a one-dimensional spin–orbit-coupled Bose–Einstein condensate (SOC-BEC) 
across the phase transition from plane-wave (PW) to stripe (ST), incorporating phenomenological damping. 
In the dissipation-free case, a state stagnation phenomenon emerges during the PW–ST quench: for slow quenches, 
the system remains trapped in the PW phase due to the energy gap induced by critical slowing down, 
which prevents spontaneous relaxation to the stripe ground state. 
To explore this phenomenon and examine the universal scaling predicted by the Kibble–Zurek mechanism (KZM) in open systems, 
we introduce a dissipative Gross–Pitaevskii equation with a phenomenological damping term. 
Numerical simulations reveal that weak dissipation preserves the expected KZM power-law scaling for the freeze-out time and defect density, 
whereas strong dissipation or long quench times lead to significant deviations. 
Our results demonstrate that the KZM remains applicable in dissipative quantum systems under appropriate conditions, 
providing insights into nonequilibrium dynamics in open quantum systems.
\end{abstract}

\maketitle
\section{Introduction}
The Kibble-Zurek mechanism (KZM) provides a universal framework for describing non-equilibrium dynamics across continuous phase transitions. 
Originally proposed in the context of cosmology and later extended to condensed matter systems \cite{1,2,3,4,5,6}, 
KZM predicts the scaling of topological defect formation as a system is driven through a critical point. In recent years, 
ultracold atomic gases, particularly Bose-Einstein condensates (BECs), have emerged as ideal platforms for experimentally testing KZM predictions 
due to their high controllability and coherence \cite{7,8,9,10}. To date, 
KZM scaling law has been successfully observed in BEC systems undergoing both continuous \cite{12,9b,9c} 
and first-order \cite{19a,20} phase transitions, confirming the robustness of this mechanism.

A particularly important system for studying critial phenomena is the spin-orbit-coupled Bose-Einstein condensate (SOC-BEC), 
which exhibits multiple quantum phases including the plane-wave (PW) phase and the stripe (ST) phase, 
the latter being regarded as a supersolid due to its spontaneously broken spatial translational symmetry \cite{9,9a}. 
In a recent study, we systematically investigated the quench dynamics between these two phases and observed asymmetric behavior: 
while the ST-to-PW transition followed KZM predictions, the reverse PW-to-ST quench exhibited a striking state stagnation effect, 
where the system remained trapped in the PW phase even after the quench was completed \cite{20}.
The mechanism of state stagnation remains an unresolved problem, thus will be further researched in this article.

The KZM has successed in closed quantum systems described by Hamiltonian evolution.
However, real-world implementations inevitably involve dissipation. 
In cold-atom experiments, factors such as Landau damping, continuous evaporation, 
and coupling to thermal reservoirs introduce unavoidable dissipative effects \cite{13,15,19}. 
Recent theoretical and experimental efforts have extended KZM to open quantum systems \cite{ad2,ad3,ad4}.
These work revealed that dissipation can modify critical exponents and 
confirmed KZM scaling can remain valid under the condition of weak dissipation.

Despite the growing body of research on KZM in open quantum systems, 
the specific conditions under which the scaling law holds remain system-dependent and require case-by-case analysis. 
In this work, we explore this question within a one-dimensional spin-orbit-coupled Bose-Einstein condensate across the plane-wave to stripe phase transition—
a system in which the absence of dissipation results in a pronounced state stagnation effect. 
By introducing a phenomenological damping term, we aim to establish a quantitative criterion for the validity of KZM scaling in the presence of dissipation, 
and explore how dissipation suppresses the state stagnation and enables the system to relax into the stripe ground state. 
Our results provide a concrete framework for understanding non-equilibrium dynamics in dissipative quantum systems, 
and offer guidance for future experimental realizations in ultracold atomic gases.

\section{Method and Model}
\subsection{Kibble-Zurek mechanism}
Consider a homogeneous system undergoing a continuous phase transition, its correlation length $\xi$ and correlation time $\tau$ will diverge as:
\begin{equation}\label{}
  \xi=\xi_0 |\epsilon|^{-\nu},\, \tau=\tau_0 |\epsilon|^{-\nu z}.
\end{equation}
A parameter $\epsilon$ is introduced to describe the phase transition distance.
$\nu$ and $z$ are critical exponents determined by the universal class of phase transition.
$\xi_0$ and $\tau_0$ are two constants.

The diverging correlation time implies that no matter how slowly the system is driven across the transition, its evolution cannot be
adiabatic when close to the critical point.
According to the KZM \cite{3}, when the evolution time approaches a frozen time $\hat{t}$, the evolution will become nonadiabatic,
and $\hat{t}$ is obtained by
\begin{equation}\label{}
  \tau(\hat{t})=\epsilon(\hat{t})/\dot{\epsilon}(\hat{t}).
\end{equation}
For a linear quench that $\epsilon=t/\tau_q$, where $\tau_q$ is the quench time, we have
\begin{equation}\label{hat}
 \hat{t}\propto \tau_q^{\frac{\nu z}{1+\nu z}},
\end{equation}
and the correlation length that freezes in at $\hat{t}$ is
\begin{equation}\label{}
 \hat{\xi}\propto\tau_q^{\frac{\nu}{1+\nu z}}.
\end{equation}
As a result, the defect density for a $d$-dimension system is given by
\begin{equation}\label{}
  n_{kz} \propto\hat{\xi}^{-d} \propto \tau_q^{-\frac{d \nu}{1+\nu z}}.
\end{equation}

\begin{figure}
 \centering
\includegraphics[width=8.8cm]{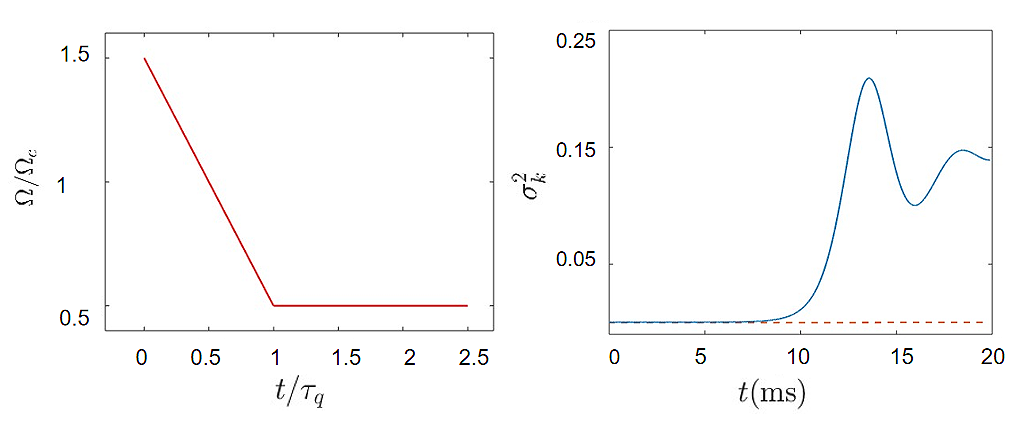}
 \caption{\label{omt} (a) Rabi frequency $\Omega$ as a function of time during the quench process.
 (b) Quench process from PW to ST without dissipation, showing the system's order parameter variation over time.
 Blue line corresponds to the fast quench scheme ($\tau_q=0.4$ ms), and red line corresponds to the slow quench scheme ($\tau_q=20$ ms).}
\end{figure}

\begin{figure}
 \centering
\includegraphics[width=8.5cm, height=7cm]{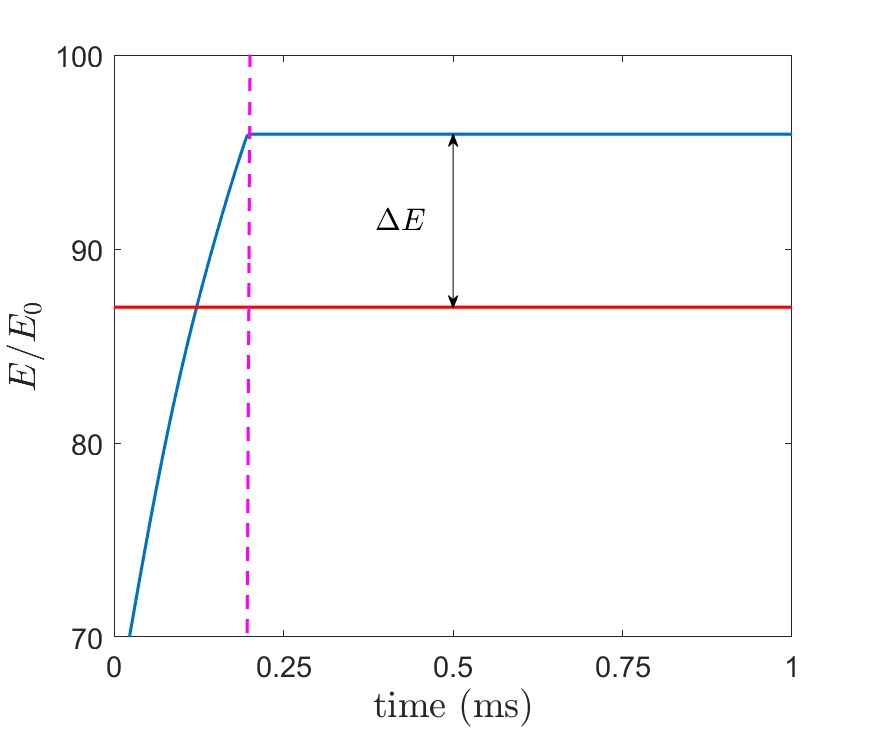}
 \caption{\label{egt} Real-time evolution curve of the system's energy during the PW-ST quench process in non-dissipative scheme.
 The purple horizontal line indicates the ground state energy of the Hamiltonian at the final moment, 
 and the red dashed line denotes the time at which the quench process ends.}
\end{figure}

\begin{figure*}[ht]
 \centering
\includegraphics[width=16cm]{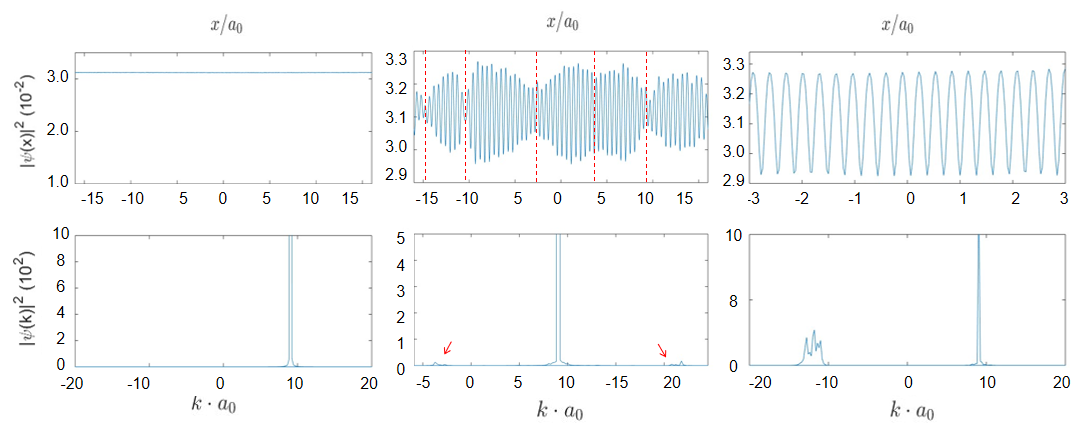}
 \caption{\label{fgps} Position distribution and momentum distribution at the final state.
 (Left) Non-dissipative scheme, slow quench (20ms), the system has no stripes in position space and remain at a single momentum state.
 (Middle) Non-dissipative scheme, fast quench (0.4ms), showing slight excitation on both sides of the main momentum peak.
 A domain wall can be identified as a consecutive minima of the envelope of wavefunction, and marked by the red dash line.
 (Right) Dissipation scheme, slow quench (20ms),the final state has denser stripes and significant momentum bifurcation.
}
\end{figure*}

\begin{figure}[htb]
 \centering
\includegraphics[width=8.2cm]{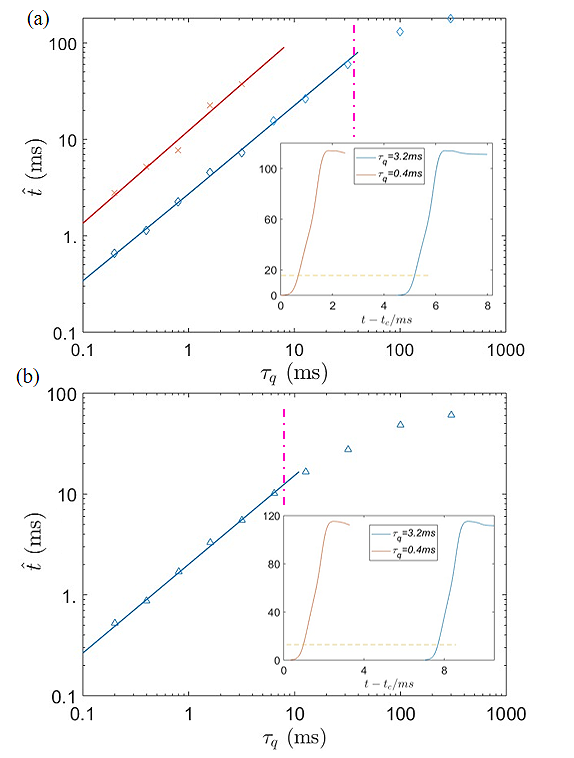}
 \caption{\label{fgtd} Scaling law of freeze-out time, for $\gamma=0.01$ (a) and $\gamma=0.1$ (b).
 The red mark in the left pannel corresponds to the non-dissipative scheme, with the fitted curve $\ln(\hat{t})=0.90\ln(\tau_q)-1.6$.
 The blue mark correspond to the dissipation scheme, with the fitted curve (a) $\ln(\hat{t})=0.89\ln(\tau_q)+1.1$; 
 (b) $\ln(\hat{t})=0.87\ln(\tau_q)+0.7$. 
 The vertical pink lines mark the threshold quench times beyond which the power-law scaling no longer holds. 
 Inset: Evolution of the order parameter $\sigma_k^2$ over time during the quench.}
\end{figure}

\begin{figure}[htb]
 \centering
\includegraphics[width=8cm]{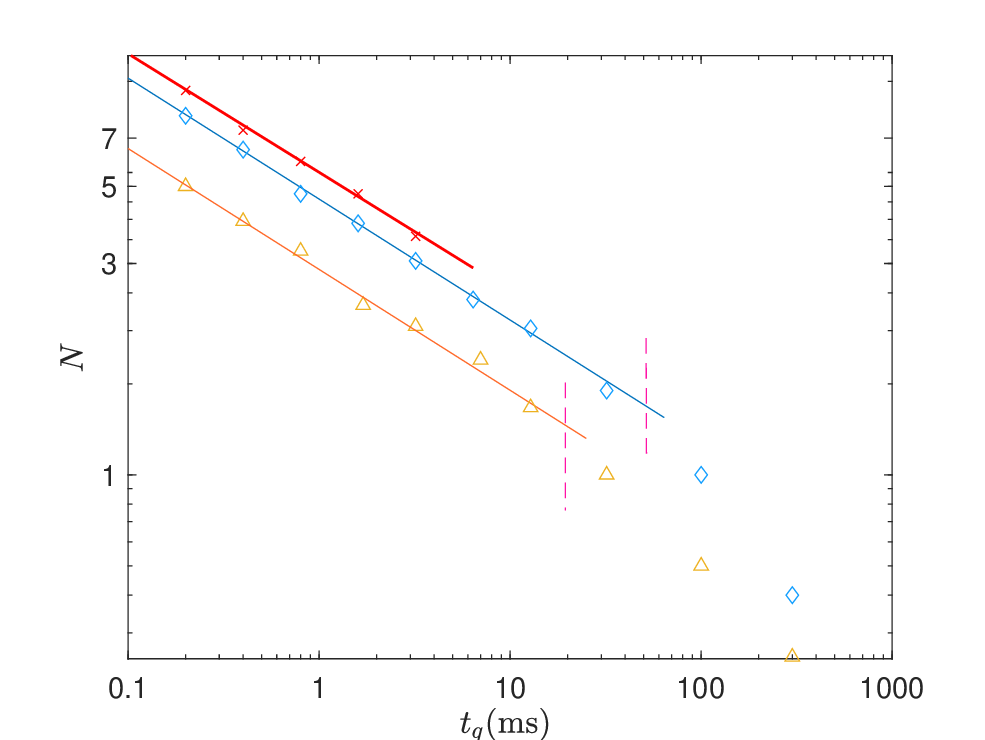}
 \caption{\label{fgnq} Scaling law of density of defects.
 The red fork corresponds to the non-dissipative scheme, with the fitted curve $\ln(N_q)=-0.50\ln(\tau_q)+3.6$.
 The blue diamond corresponds to the dissipation scheme for $\gamma=0.01$, with the fitted curve $\ln(N_q)=-0.48\ln(\tau_q)+2.1$.
 The yellow triangle corresponds to the dissipation scheme for $\gamma=0.1$, with the fitted curve $\ln(N_q)=-0.45\ln(\tau_q)+1.8$.
The vertical pink lines mark the threshold quench times beyond which the power-law scaling no longer holds.}
\end{figure}

\subsection{Spin-Orbit-Coupled Bose-Einstein Condensate}
We consider the Hamiltonian of a one-dimensional Raman-type SOC-BEC system.
Choosing $a_0 = 6$ \textmd{$\mu$m}, $t_0=\frac{a_0^2m}{\hbar}$, $E_0=\frac{\hbar^2}{a_0^2m}$ as the units of length, time and energy, respectively,
where $m$ is the mass of atom $^{87}$Rb and $\hbar$ is reduced Planck constant.
The dimensionless time-dependent Gross-Pitaevskii
equation is written as $i\partial_t\Psi = H\Psi$, where $\Psi=(\psi_1,\psi_2)^T$ is the two-component wavefunction,
and the effective Hamiltonian of the system is
\begin{eqnarray}
  H&=&\frac{(k-k_0\sigma_z)^2}{2}+\frac{\Omega}{2}\sigma_x+V_{int}, \notag \\
  V_{int}&=&\begin{bmatrix} g_{11}|\psi_1|^2+g_{12}|\psi_2|^2 & 0 \\ 0 & g_{21}|\psi_1|^2+g_{22}|\psi_2|^2 \end{bmatrix},
\end{eqnarray}
where $k_0$ is the strength of the spin-orbit coupling determined by the momentum transfer of the two Raman lasers,
$\Omega$ is the Raman coupling strength, and $\sigma_x$, $\sigma_z$ are components of the Pauli matrices.

The interaction between the spin components are described by coupling constants $g_{ij}=\frac{4\pi a_{ij}}{a_0}$, 
with $a_{ij}$ being the scattering length.
In the following, we assume that $g_{11}=g_{22}=g$, and $g_{12}=g_{21}$.

This system has three phases: zero momentum, plane-wave phase, and stripe phase \cite{9a}. 
This work focuses on the plane-wave and stripe phases with the following expressions:

(i) Stripe (SP) phase: it is a superposition of the momentum states of $k$ and $-k$,
\begin{equation}\label{}
  \psi_s = \sqrt{\frac{n}{2}}[\begin{pmatrix} \cos{\theta}\\-\sin{\theta}\end{pmatrix} e^{i(kx+\phi)}
   + \begin{pmatrix} \sin{\theta}\\-\cos{\theta}\end{pmatrix} e^{-ikx}],
\end{equation}
where $\phi$ is the relative phase of $C_1$ and $C_2$.

(ii) Plane-wave (PW) phase: the system is in one of the following states,
\begin{equation}\label{}
  \psi_\uparrow = \sqrt{n}\begin{pmatrix} \cos{\theta}\\-\sin{\theta}\end{pmatrix}e^{ikx},\,
  \psi_\downarrow = \sqrt{n}\begin{pmatrix} \sin{\theta}\\-\cos{\theta}\end{pmatrix}e^{-ikx}.
\end{equation}

Here we introduce the parameter settings of the system.
We apply the periodic boundary condition for our system, and the period of the condensate is set as $192$ \textmd{$\mu$m}.
The intra-species scattering lengths are $a_{11}=a_{22}=110a_r$, and the inter-species scattering lengths are $a_{12}=a_{21}=100a_r$, 
where $a_r$ is the Bohr radius.
SOC strength is $k_0=\sqrt{2}\pi/\lambda$, where $\lambda=784 \textrm{nm}$ is the wavelength of the Raman laser.
In addition, we set the number of atoms $N=3.2\times10^5$,
i.e. the atomic density is smaller than the critical density $n_c$ ($n_c=2.14\times10^{10} \textmd{m}^{-1}$).
When we manipulate the phase transition parameters, we adopt a linear quench scheme, expressed as follows:
\begin{equation}\label{}
 \Omega(t)=\frac{\Omega_c}{\tau_q}(t-t_c)+\Omega_c.
\end{equation}
Use $\Omega_i$ as the initial Raman frequency, then $t_c$ can be calculated by
\begin{equation}\label{}
  t_c=\frac{\Omega_c-\Omega_i}{\Omega_c}\tau_q.
\end{equation}
We quench the parameter $\Omega_i$ from to $\Omega_f$ in a time $\tau_q$, and then keep it unchanged, as illustrated in Fig.\ref{omt}.

For the SP-PW phase transition, the variation of momentum can be chosen as the order parameter:
$\sigma_k^2=\int k^2 n(k)dk/\int n(k)dk-(\int kn(k)dk/\int n(k)dk)^2$.
For the standard plane-wave state, $\sigma_k^2=0$;
for the stripe state, $\sigma_k^2=k_s^2$.
By tracking the growth of the order parameter, we can determine when the system enters the new phase.
To begin the simulation, we first perform an imaginary-time evolution with $\Omega_i$ to get an initial ground state $\psi_0(x)$.
Then we add a noise term to simulate the fluctuation of the environment,
\begin{equation}\label{}
  \psi(x,t_0) = \psi_0(x) + \Delta \psi(x),
\end{equation}
\begin{equation}\label{}
  \Delta \psi(x) = \begin{pmatrix} \alpha^1(x)\sqrt{n_1(x)}\\ \alpha^2(x)\sqrt{n_2(x)} \end{pmatrix}.
\end{equation}
Here $n_i(x)=|\psi_i(x)|^2$ is the density of component $i$, and $\{\alpha^m(x)\}$ are independent complex Gaussian random variables with $\langle\alpha^m(x)\alpha^n(x')\rangle=\frac{1}{2}\delta_{mn}\delta_{xx'}$.

The wave function after the quenching process can be calculated as
\begin{equation}\label{}
  \psi(x,t_f) = \hat{T}e^{-i\int ^{t_f}_{t_0}Hdt} \, \psi(x,t_0). 
\end{equation}

\section{The state stagnation phenomenon}
We numerically simulated the GPE and tracked the evolution of the parameter $\sigma_k^2$, 
with the results shown in Fig.\ref{omt}.
For the slow quench scenario ($\tau_q=20$ ms), $\sigma_k^2$ remains zero, 
and the system stays in the metastable plane-wave phase. 
For the fast quench scenario ($\tau_q=0.4$ ms), however, 
rises rapidly after crossing the phase transition point and eventually stabilizes at a non-zero value, 
indicating successful formation of the stripe phase.
Furthermore, the saturated value of the order parameter is considerably smaller than the theoretical expectation for a perfect stripe phase.
This indicates that the final state is not a true stripe ground state but rather an excited state bearing stripe-like modulations, 
a manifestation of the residual momentum excitations injected during the fast quench.

Next, we analyze the cause of the system's final state from the perspective of energy evolution.
When changing the Rabi frequency parameter to cross the phase transition point,
critical slowing down mechanism indicates that the system's state cannot adiabatically follow the Hamiltonian,
thus freezing at a certain time $\hat{t}$, whose expression is determined by Eq.(\ref{hat}).
After the quench process ends, the energy of the system can be estimated as:
\begin{equation}\label{}
  E = \langle \psi(\tau_q)|H(\Omega_f)| \psi(\tau_q)\rangle
\end{equation}
It is clearly greater than the corresponding ground state energy $E_g$.
Therefore, in the absence of energy dissipation, the system cannot naturally evolve to reach the new ground state,
creating the phenomenon of an infinite relaxation time.
We recorded the changes in system energy during the evolution in Fig.\ref{egt}, which confirms the above reasoning.

The energy gap caused by critical slowing down during a phase transition is a common phenomenon.
However, whether state stagnation appears depends on the size of the energy gap.
After the quench ends, the system might also evolve into a defective ground state,
where the energy difference between the final state and the ground state compensates for the defect energy (the energy cost to form domain walls).
For example, in our previous study \cite{20}, we simulated the quench process from ST to PW and did not observe the state stagnation,
precisely because the energy gap was relatively small.

\section{The scaling law under the dissipation scheme}
For the PW-ST quench process in the dissipation-free case, the order parameter exhibits no detectable change during slow quenches ($\tau_q > 10$ms),
indicating that the system remains trapped in the metastable PW phase. 
Verifying the universal scaling predicted by the KZM therefore requires accessing a broader range of quench times $\tau_q$ to capture the power-law behavior. 
On the other hand, practical cold-atom implementations inevitably involve dissipative processes. 
These include, for instance, inelastic collisions between condensed atoms and thermal components, as well as continuous atom loss and heating effects \cite{19}. 
Such dissipative channels are intrinsic to open systems. 
Therefore, a thorough examination of whether and how the KZM scaling survives in the presence of dissipation is crucial for 
validating its universality under realistic experimental conditions.
\begin{equation}\label{damp}
  i\partial_t\Psi = (1-i\gamma)(H-\mu)\Psi,
\end{equation}
where $\gamma > 0$ is the dissipation rate, and $\mu$ is the chemical potential.
The exact value for the damping parameter is expected to depend on a variety of factors.
For example in BEC system, $\gamma$ represents the rate at which the excited components turn into the condensate. 
This may be approximated by the transition probability that was estimated in a work by Gardiner et al. \cite{ad1} 
using the quantum kinetic theory.

We performed numerical simulation evolution using Eq.(\ref{damp}) and 
plotted the density distribution and momentum distribution of the final state, as shown in Fig.\ref{fgps}. 
Compare the non-dissipative scheme with the same quench time, 
the dissipation dynamic obtains a final state at stripe phase.
This indicates the disappearance of the state stagnation phenomenon in the dissipation system.

Next, we explore the scaling relation of the freeze-out time with respect to the quench time.
Considering the possible influence of the dissipation rate on the scaling law,
we conducted tests for $\gamma=0.01$ and $\gamma=0.1$, with results shown in Fig.\ref{fgtd}.
It can be found that the relaxation time $\hat{t}$ in dissipative case is shorter than that in non-dissipative case.
This is understandable, when crossing the critical point, 
dissipation reduce the energy of system and make the state follow the ground state more synchronously. 
On the other hand, after excluding the data on the right side of the Fig.\ref{fgtd} which deviated from the power law, 
we obtained the combined critical exponent as 
$\frac{\nu z}{1+\nu z}=0.88$ (with $\gamma=0.01$), $\frac{\nu z}{1+\nu z}=0.86$ (with $\gamma=0.1$).
That is very close to the result in the non-dissipative case, thus verifying the universality of KZM.

It is worth noting that the influence of dissipation on the KZM scaling is primarily governed by the dimensionless parameter $\gamma \tau_q / t_0$. 
Our numerical results indicate that for sufficiently weak dissipation satisfying $\gamma \ll t_0/\tau_q$ (i.e., $\gamma \tau_q / t_0 \ll 1$),
the power-law scaling remains approximately valid. Conversely, when $\gamma \tau_q / t_0 \gtrsim 0.02$, 
a significant deviation from the KZM power law is observed in the freeze-out time (see Fig.\ref{fgtd}). 
This threshold is obtained phenomenologically from our simulations and may be system-dependent.

To identify defects in the stripe phase, we extract the envelope of the density profile using a Hilbert transform. 
The resulting envelope exhibits local minima corresponding to the nodes of the density modulation. 
A defect is identified as the region between two consecutive envelope minima where the spacing deviates from the ideal value by more than a preset threshold. 
This method effectively counts phase slips in the density modulation and provides a robust measure of the defect density.

Subsequently, we investigate the scaling of the defect density with respect to the quench time $\tau_q$; the results are presented in Fig.~\ref{fgnq}. For sufficiently small $\tau_q$, the power-law scaling predicted by the KZM remains valid, and the extracted combined critical exponent is close to that obtained in the dissipation-free case. In contrast, when $\tau_q \sim t_0 / \gamma$, the defect density significantly decreases and deviates from the power-law behavior. This reduction is attributed to defect merging induced by continuous energy dissipation, which results in a lower defect count in the dissipative system compared to the non-dissipative case.

\section{Discussion and conclusion}
In this work, we have investigated the quench dynamics across the plane-wave to stripe phase transition 
in a one-dimensional spin-orbit-coupled Bose-Einstein condensate under phenomenological dissipation. 
In the non-dissipative case, slow quenches lead to a state stagnation effect where the system becomes trapped in a PW phase, 
preventing trsnstion to the stripe ground state. 
By introducing a damping term, we find that the Kibble–Zurek mechanism remains applicable in open systems under weak dissipation, 
with the freeze-out time and defect density following the expected power-law scaling. 
However, strong dissipation or slow quenches cause significant deviations due to enhanced relaxation and defect merge. 
Our results establish the robustness of KZM in dissipative quantum systems under appropriate conditions and 
provide a foundation for exploring non-equilibrium dynamics in cold-atom experiments.

\begin{acknowledgments}
This work is supported by the National Natural Science Foundation of China (No.92065113),
Innovation Program for Quantum Science and Technology (No.2021ZD0301201),
the Fundamental Research Funds for the Central Universities (WK2030000088),
and the University Synergy Innovation Program of Anhui Province (No.GXXT-2022-039).
\end{acknowledgments}

\bibliography{ref.bib}
\end{document}